\documentstyle[12pt,epsf]{article}
\input epsf.tex
\catcode`\@=11

\begin{document}
\begin{titlepage}
\samepage{
\setcounter{page}{1}
\rightline{FERMILAB-PUB-99/242-T}
\rightline{OSU-HEP-99-04}
\vfill
\begin{center}
{\Large \bf Asymmetrical Large Extra Dimensions\\}
\vspace{.3in}
 {\large
Joseph Lykken$^{(a)}\footnote{e-mail: lykken@fnal.gov}$ and
 Satyanarayan Nandi$^{(a,b)}\footnote{Summer visitor at Fermilab,
e-mail: shaown@okstate.edu}$\\}
\vspace{.25in}
{\it $^{(a)}$Theory Dept., Fermi National Accelerator Laboratory,
     Batavia, IL 60510.\\[3mm]}
{\it $^{(b)}$Department of Physics, Oklahoma State University, 
     Stillwater, OK 74078.\\}
\end{center}
\vfill
\vfill
\begin{abstract}
We study scenarios in which there is a hierarchy of two sets of
large compactified extra dimensions. One particularly interesting
case has a single millimeter size extra dimension and five TeV$^{-1}$ size
dimensions. The Standard Model gauge bosons have Kaluza-Klein
excitations with respect to one of the TeV scale dimensions. We
discuss astrophysical constraints on this scenario, as well
as prospects for signals at future high energy colliders.
\end{abstract}
\vfill}
\end{titlepage}


\catcode`@=11
\long\def\@caption#1[#2]#3{\par\addcontentsline{\csname
  ext@#1\endcsname}{#1}{\protect\numberline{\csname
  the#1\endcsname}{\ignorespaces #2}}\begingroup
    \small
    \@parboxrestore
    \@makecaption{\csname fnum@#1\endcsname}{\ignorespaces #3}\par
  \endgroup}
\catcode`@=12


\newcommand{ \slashchar }[1]{\setbox0=\hbox{$#1$}   
   \dimen0=\wd0                                     
   \setbox1=\hbox{/} \dimen1=\wd1                   
   \ifdim\dimen0>\dimen1                            
      \rlap{\hbox to \dimen0{\hfil/\hfil}}          
      #1                                            
   \else                                            
      \rlap{\hbox to \dimen1{\hfil$#1$\hfil}}       
      /                                             
   \fi}                                             %


\newcommand{\hmu}{{\hat\mu}}
\newcommand{\hnu}{{\hat\nu}}
\newcommand{\hrho}{{\hat\rho}}
\newcommand{\hh}{{\hat{h}}}
\newcommand{\hg}{{\hat{g}}}
\newcommand{\hk}{{\hat\kappa}}
\newcommand{\tA}{{\widetilde{A}}}
\newcommand{\tP}{{\widetilde{P}}}
\newcommand{\tF}{{\widetilde{F}}}
\newcommand{\th}{{\widetilde{h}}}
\newcommand{\tp}{{\widetilde\phi}}
\newcommand{\tchi}{{\widetilde\chi}}
\newcommand{\te}{{\widetilde\eta}}
\newcommand{\vn}{{\vec{n}}}
\newcommand{\vm}{{\vec{m}}}
\newcommand{\gsim}{\lower.7ex\hbox{$\;\stackrel{\textstyle>}{\sim}\;$}}
\newcommand{\lsim}{\lower.7ex\hbox{$\;\stackrel{\textstyle<}{\sim}\;$}}


\section{Introduction} \label{sec:intro}

New developments
in superstring theory \cite{polchinski} have led to a
radical rethinking of the possible phenomenological implications
of the existence of extra spatial dimensions.
In superstring theory there are regions of moduli space
where compactification radii become large while the string
coupling, gauge couplings, and Newton's constant remain
fixed \cite{witten,lykken}.
The scale $R$ of these large extra dimensions determines the relation
\cite{dimopoulos}
between the usual Planck mass $M_P$ and an effective higher dimensional
Planck scale $M_*$:
\begin{equation}
M_P^2 = M_*^{n+2}\,R^n\ ,
\label{mas}
\end{equation}
where for the moment we have taken all $n$ large extra dimensions to
have the same size $R$. The scale $M_*$ is related to the string scale
$M_S$ in a way which depends in general on the vacuum values of other
moduli, including the size of other (smaller) extra dimensions.
Roughly speaking, $M_S$ plays the role of the ultraviolet
cutoff $\Lambda$ for the effective (nonrenormalizable) Kaluza-Klein theory.

Recently it was observed \cite{dimopoulos}
that this scenario can be phenomenologically viable for $n\ge2$, if we
assume that the fields of the Standard Model are confined
to a three-dimensional brane or intersection of branes in
the larger dimensional space. Assuming further that the scale
of the brane tension is of the order of the cutoff $\Lambda$
or larger, the resulting effective theory consists of 
$(3+1)$-dimensional Standard Model fields coupled to $4+n$ gravity
and, perhaps, other $(4+n)$-dimensional ``bulk'' fields.
With these assumptions the phenomenological constraints from
gravity experiments, collider physics, and astrophysics are
surprisingly weak \cite{dimopoulos}--\cite{hall},
allowing $1/R$ scales as low as 10 MeV to $10^{-3}$ eV, for cutoff
scales $\Lambda$ in the range $1-100$ TeV.

It has also been shown in superstring theory that it is
possible to obtain $d=4$ N=1 supersymmetric chiral gauge
theories confined to the world-volumes of stable configurations
of intersecting D-branes \cite{chiral}. The regions of string moduli
space where such configurations have a perturbative description
is not necessarily incompatible with the region where large
extra dimensions may occur.
Thus within our current knowledge (or ignorance) of superstrings 
it is not implausible to imagine that the Standard Model is
confined to a brane configuration 
\cite{past,Ibanez:1997rf,Antoniadis:1998ig,Shiu:1998pa},
while large compactified
dimensions are probed only by gravity and other bulk
fields \cite{dimopoulos,tye}. 

It is equally possible that the Standard Model gauge theory is confined
to a brane with more than three spatial dimensions, with the brane
wrapped around one or more extra large dimensions. In this case collider
limits constrain the compactification scale to be larger
than about 1 TeV \cite{ant,antben}. The Standard Model gauge couplings
will exhibit power law running above the compactification scale at which
the gauge bosons begin to probe one or more extra dimensions \cite{DDG}.
If we equate the gauge coupling unification roughly with the string scale,
this would imply that the string scale is no more than one or two orders
of magnitude above the compactification scale. This scenario is not, however,
compatible with the simple scaling relation Eq.~(\ref{mas}), if we assume
that no extra dimensions are larger than an inverse TeV. Thus it has been
generally assumed that the ``millimeter'' and ``TeV'' large extra dimension
pictures are mutually exclusive.

The basic scaling relation Eq.~(\ref{mas}) is the simplest one possible,
and many researchers have observed that a variety of asymmetrical scenarios
are also possible.
With a total of 7 extra compactified dimensions at a generic
point in superstring/ M theory moduli space, one can hypothesize the
existence of several separate compactification scales, all distinct
from $M_S$. In this paper we will consider the next simplest case,
where there is a hierarchy between two sets of ``large'' extra dimensions.
Thus:
\begin{eqnarray}
M_P^2 &= M_*^{n+2}\,R^n\ ,\\
&= M^{n+m+2}\,R^n\,r^m\ ,
\label{newmas}
\end{eqnarray}
where $n$$+$$m\le 7$, and we now call $R$ the size of the ``very large''
extra dimensions, while $r$ denotes the size of the merely ``large''
extra dimensions.

At this point we must decide which large dimensions, if any, are probed by
Standard
Model particles. We could assume that only the graviton and the right--handed
neutrino probe any of the $n$$+$$m$ extra dimensions. In this case the
asymmetrical setup is useful for evading the rather stringent astrophysical
bounds \cite{dimopoulos} on the case $n$$=$$2$ in the
symmetrical
scenario. One also can obtain an attractive scenario for neutrino masses
\cite{smirnov}. 

Another possibility for these more general scenarios is that the brane
volume containing the Standard Model is transverse to the very large
dimensions of size $R$, but does extend in one or more of the large extra
dimensions of size $r$. This has dramatic consequences for the evolution
of the Standard Model gauge couplings, as noted earlier \cite{DDG}. Above the
energy threshold $1/r$, the logarithmic running of the gauge couplings will
be replaced by (effectively) power law running; nevertheless gauge coupling
unification may still occur, albeit at a rather lower energy scale.
For example, we can assume that Standard Model chiral matter is confined to
a 3-brane volume, while the gauge and Higgs fields are confined to a 4-brane
with one dimension compactified at scale $r$. Then the analysis of \cite{DDG}
indicates that gauge coupling unification can still occur, at a scale which is
no more than about 20 times $1/r$.

Roughly speaking, we ought to identify the gauge coupling unification
scale with the string scale $M_S$. Thus the scale $1/r$ is naturally
one order of magnitude less than $M_S$. However, we ought to allow ourselves
at least one additional order of magnitude of stretch in this ratio.
This is because extra matter with Standard Model charges can effect
the running of the gauge couplings, and there may be large threshold effects
at the string scale. On the other hand, it is certainly more difficult to
arrange a hierarchy between the scale $1/r$ and the string scale if the
Standard Model particles probe more than one extra dimension.
New interpretations of gauge coupling unification also need
to be considered \cite{ibanez,arkani}.

A particularly dramatic case of this class of asymmetric scenarios occurs when
$n=1$, $m=5$. In this case (\ref{newmas}) becomes:
\begin{eqnarray}
M_P^2 &= M_*^3\,R\ ,\\
&= M^8\,R\,r^5\ .
\label{ourmas}
\end{eqnarray}
We can assume that the compactification scale $1/R$ is $10^{-3}$ eV, which is 
in the
range accessible to millimeter gravity experiments, as well as providing the
appropriate neutrino mass scale for an explanation solar neutrino 
data \cite{smirnov}. For the compactification
scale $1/r$, we choose 1 TeV, attempting to saturate the current collider
lower bounds. With these inputs (\ref{ourmas}) implies:
\begin{eqnarray}
&M = 137 {\rm\ TeV}\ ,\\
&M_* = \left({M\over 1 {\rm\ TeV}}\right)^{8/3} = 5 \times 10^5 {\rm\ TeV}\ .
\label{scales}
\end{eqnarray}
Note that $M\sim 100$ TeV is compatible with our argument above that
we expect $M_S$ to be not more than two orders of magnitude above $1/r$.
In this scenario both $M_P$ and $M_*$ are parameters measuring the strength
of gravitational couplings; they do not correspond to scales of new
ultraviolet physics. $M$, on the other hand, can be regarded as roughly
equal to the ultraviolet cutoff $\Lambda$.

This scenario is quite novel in that it predicts a single millimeter size
extra dimension, as opposed to two such dimensions in the scenario
based on (\ref{mas}). It also predicts TeV scale Kaluza-Klein (KK) thresholds
for the Standard Model gauge bosons (and perhaps the Higgs), but probably not
for
quarks and leptons. Thus this scenario is constrained both by
phenomenological considerations similar to those discussed in
\cite{dimopoulos} for millimeter size extra dimensions, as well as
those discussed in \cite{antben} for TeV scale Kaluza-Klein gauge bosons.

\section{Astrophysical constraints}

Here we review the constraints on a single mm size extra dimension,
following the discussion of Arkani-Hamed et al \cite{dimopoulos} and later
papers \cite{barger,perelstein,hall}. Note
that in our scenario the effective ultraviolet cutoff for Kaluza-Klein
mode sums is $M \sim 100$ TeV. At energy scales $\sqrt{s}$ less than $1/r$, the
effects of KK graviton emission are greatly suppressed in our scenario.
Cross sections for real emission of KK gravitons scale like
\begin{equation}
\label{sf}
\sigma \sim {\sqrt{s}\over M_*^3}\; .
\end{equation}
Above the scale $1/r$ but below the scale $M$, KK graviton emission is
less suppressed but still very soft; cross sections scale like:
\begin{equation}
\label{soft}
\sigma \sim {s^3\over M^8}\; .
\end{equation}
Of course in this energy regime we also have the possibility of real emission
of KK gauge bosons (or Higgs); these cross sections are proportional to powers
of Standard Model couplings and do not have any extra suppression above 
the kinematic threshold.

Effects due to virtual KK exchanges are much less suppressed than real KK
emission in the low energy region. The cross sections scale like
$s^3/M^8$, or like $s/M^4$ in channels where there is interference with
Standard Model diagrams.

Thus we expect that the effects of real KK graviton emission are completely
negligible at
energies below $1/r$, while virtual KK graviton effects are perhaps marginally
observable.
The remainder of this section is devoted to confirming these expectations.

Most important to check are the astrophysical constraints.
Some constraints come from the fact that
the energy loss due the KK graviton emissions by the sun, red giants,
and the supernova 1987A must not exceed certain upper bounds.
The most stringent
bound comes from the supernova, where the dominant process for the energy loss
is
nucleon--nucleon bremstrahlung (N+N$\rightarrow$ N+N+graviton). Following
\cite{perelstein}, the energy loss per gram per second for our scenario is
given by
\begin{equation}
\label{nnb}
\stackrel{\cdot}{\varepsilon}=3.9\times 10^{22} {\rm erg}\,
g^{-1}s^{-1}\left(X_{n}^2+X_{p}^2+
4.3 X_{n} X{p}\right)\rho_{14}T_{\rm MeV}^{4.5} M_*^{-3}
\end{equation}
where$ X_{n}, X_{p}$ are the neutron and proton fractions, $\rho_{14}$ is
the density measured in units of $10^{14}$ grams per $cm^{-3}$,
$T_{\rm MeV}=T_{SN}/$1 MeV, and $M_*$ is in TeV units.
Using $T_{\rm MeV}=30, X_{n}=0, \rho_{14}=3$, and
requiring that the energy loss rate to gravitons, as given 
by Eq.\,\ref{nnb} , do not exceed $10^{19}\, {\rm erg} \,g^{-1} s^{-1}$,
we obtain,
\begin{equation}
M_* \geq 3700 {\rm\ TeV},\quad  \rightarrow M \geq 22 {\rm\ TeV}\;.
\end{equation}

We now consider the cosmological bound arising from the absence of ``MeV
bumps" in the cosmic diffuse gamma (CDG) radiation background as set by the
recent measurement using COMPTEL instrument, in the energy range of 0.8 to
30 MeV \cite{sckap} . Neutrino-antineutrino,
and photon-photon annihilation will produce
massive KK gravitons, and their subsequnt decays to two photons will produce
such a bump. Recent measurement of the photon spectrum is fitted well by a 
continuous distribution,
\begin{equation}
{dn_{\gamma}\over{dE}} = A  \left({E\over{E_{0}}}\right)^a
\end{equation}
where $A=\left(1.05\pm 0.2\right)\times
10^{-4}$ MeV$^{-1}$ $cm^{-2}s^{-1} ster^{-1}$,
$E_{0}$$=$5 MeV, $a$$=$$-2.4\pm0.2$.
Following the work of \cite{hall}, the contribution to the energy
distribution from the annihilation of two photons, and three flavors
of neutrino-antineutrinos are given by
\begin{equation}
\label{cdg}
\left({dn_{\gamma}\over{dE}}\right)_{T_{*}=1{\rm\ MeV}} = 7 \alpha
_{n}\left(E\right)
\left({M\over\rm TeV}\right)^{-\left(n+2\right)}
{\rm MeV}^{-1} {cm}^{-2} s^{-1}
{ster}^{-1}.
\end{equation}
where
\begin{equation}
 \alpha _{n}\left(E\right) = 4.6\times {10}^{-6\left(n-2\right)} 
{{2\pi^{n/2}}\over{\Gamma\left(n/2\right)}}
{f_n(E,T_*=1{\rm\ MeV})\over ({\rm MeV})^{n+5/2}}
{\left(E\over MeV\right)^{\frac{1}{2}}} \; .
\end{equation}
with the expression for $f_n$ given in \cite{hall}.
In our scenario this becomes

\begin{equation}
\left({dn_{\gamma}\over{dE}}\right)_{T_{*}=1{\rm\ MeV}}
= 7 \alpha _{1}\left(E\right)\left({M_*\over\rm TeV}\right)^{-3}
cm^{-2}s^{-1} ster^{-1}.
\end{equation}
We have calculated $\alpha_{1}\left(E\right)$ for
the energy range $E=1$ to 10 MeV. The bound on $M_*$ is given by
\begin{equation}
\label{cdgb}
\left({M_*\over\rm TeV}\right)
\geq 7 \alpha _{1}\left(E\right)
\left( \frac{\left(\frac{d \eta_\gamma}{dE}\right)_{\rm measured} }
{{\rm MeV}^{-1} cm^{-2} s^{-1} ster^{-1} }\right)^{-1}
\end{equation}
In the range of $E=1$ to 10 MeV, the bound for M as obtained from
Eq. (16) is fairly flat, varying from 24 to 48 TeV.
The most stringent bound for M comes
using E equal to 3 or 4 MeV, for which 
the value of 
$\alpha_{1}\left(E\right)$ is $1.55 \times 10^9$ and $7.1 \times 10^8$
respectively.
Using these values, we obtain from Eq. (16):

\begin{equation}
M_* \geq 30000 {\rm\ TeV},\quad \rightarrow M \geq 48 {\rm\ TeV}\;.
\end{equation}
Thus we find that our scenario evades even the extremely stringent
astrophysical
bounds on a single millimeter size extra dimension.

\section{Collider constraints}

In this section, we discuss the laboratory constraints for our model
coming from
high energy processes. In our model, the gauge and the Higgs
bosons live on a brane which contain one or more large TeV size dimensions.
Thus, their
KK excitations will contribute to the effective 4-dimensional theory.
Below the threshold $\mu_{0}=r^{-1}$
these may be observable as off-shell contributions, while in future
high energy colliders we may be able to produce some of the low lying KK states
as resonances. The interesting point to note here is that the constraints from
these
processes will give bounds on the masses of these KK states, and hence on the
scale $\mu_0$.
A number of analyses already give interesting bounds. For example
Marciano \cite{marciano} has looked into the effect 
of the $W^{*}$, a KK excitation of the W boson, to low energy weak processes,
and
obtained a bound of $ m_{W^{*}}\geq3.7$ TeV for the case of only one extra
dimension. The authors of Ref. \cite{nath} studied the effects for
the processes
$e^{+}e^{-}\rightarrow\mu^{+}\mu^{-}$ at LEP2, and the Drell-Yan process in
hadronic
colliders. They set lower bounds on $\mu_{0}$ in the range of 1 to 3
TeV.

As an example, we consider the effects of the 
KK excitations of gluons to top production at the Tevatron.
The KK excitations of the gluons will contribute to both 
$q\bar{q}\rightarrow t\bar{t}$, and $gg\rightarrow t\bar{t}$ subprocesses via
their 
exchanges in the appropriate s,t and u channels. In the light of the above
bounds,
it is unlikely that we will see such KK resonances directly in Tevatron dijet
spectra, unless the gluon KK modes turn out to be somewhat light compared to
the KK modes of the electroweak gauge bosons.

The effect of the KK
states in the $ q\bar{q}$ subprocess is to modify the propagator $1/s$ to
$D\left(s\right)$ with (ignoring the appropriate width factor)
\begin{equation}
\label{pre}
D\left(s\right)= \sum_{n=1}^{\infty}{1\over{s-{m_{n}^2}}}.
\end{equation}
with similar modification in the t and u channels. Writing
${m_{n}^2}={\mu_{0}^2}
{n^2}$ for the KK excitations of the gluons, with one extra dimension
Eq.~(\ref{pre})
reduces to (for $s\ll{\mu_{0}^2}$):
\begin{equation}
D\left(s\right)\simeq{1\over s}-{{\pi^{2}}\over{6{\mu_{0}^2}}}
\end{equation}
Thus, well below threshold, the net effect of the interference of the KK
excitations
is to reduce the subprocess cross sections by the factor $D(s)^2$.
We have calculated the cross sections for $p\bar{p}\rightarrow t\bar{t}X$ at
the 
Tevatron ($\sqrt{s}=1.8$ TeV, 2 TeV) as functions of $\mu_{0}$.
We included only the $ q\bar{q}$ contribution, since gg is small at the
Tevatron. In Fig. 1, we show the ratio $R$ defined by
\begin{equation}
R\equiv{ {\sigma_{\mu_{0}}\left(p\bar{p} \rightarrow t\bar{t}\right)}\over
{\sigma_{SM}\left(p\bar{p}\rightarrow t\bar{t}\right)}}
\end{equation}
with $m_t=175$ GeV for the Tevatron Run 2 ($\sqrt{s}=2$ TeV) as a function of
the compactification scale $\mu_{0}$ . The cross section  is expected to
be measured to about five percent accuracy at the Tevatron Run 2.
This can be used to set lower bound on the compactification
scale, $\mu_{0}$, and is expected to be around 3 TeV.
LHC experiments will raise this bound significantly, or discover
the low lying KK excitations of the gluons.

\section{New physics in hadron colliders}

The smoking gun signature of our scenario is the existence of a single
millimeter sized very large dimension, combined with dramatic new
high $p_T$ physics in future colliders. Below we discuss
some of these new physics signals relevant to hadron collider
experiments near or above the KK threshold energy $\mu_0$.

At the hadronic collider, in our scenario, the main effect will come
from $g{_1}{^*}$, the first KK excitation of the gluon which propagates
into one of the TeV$^{-1}$ scale extra dimension. In our model, the 
quarks are localized in the usual 4 space-time dimensions. Their interactions
do not conserve momenta in the extra dimensions, and hence the tree-level couplings
$\bar{q}qg_{n}^{*}$ are allowed in the effective four dimensional theory.
Since the gluons propagate in the extra dimension, fifth dimensional momentum
conservation forbids $ggg_{n}^{*}$ couplings in the effective four dimensional
theory. However, such an effective coupling is generated at the one loop level
via a quark loop similar to the gluon-gluon-Higgs coupling (ggH) in the Standard
Model. There are several new interesting phenomena in our scenario related to
these vertices, for high energy hadronic colliders, which could lead to new
signals for the proposed TeV$^{-1}$ scale extra dimenions.

\begin{enumerate}  
\item {\bf Enhancement of high $p_T$ dijet production}

The important subprocess relevant for the dijet productions are:

$gg\rightarrow g{_1}{^*}\rightarrow gg,  q\bar{q}$

$q\bar{q}\rightarrow g{_1}^{*}\rightarrow gg, q\bar{q}$

The first subprocess will be best studied at the upgraded Tevatron ($\sqrt{s}=2$
TeV), since at this energy, about 90\% of the luminosity is in the $ q\bar{q}$ channel.
At this energy, it is unlikely that we shall produce the $g{_1}{^*}$ resonance.
However, we might see an enhancement in the high $p_T$ jet cross sections due to 
the off-shell effect of $g_{1}^{*}$. This will
be similar to the single jet inclusive excess
reported by the CDF collaboration \cite{cdf};
thus, as in this case, it will important to have a firm handle on the
parton distribution functions.

At the LHC, resonance productionof the $g_{1}^{*}$ may be energetically possible.
This state will be very wide, with a width few tenths of its mass. Thus it will be 
very hard to observe the the actual resonance structure. However, since the final
 states gg , $q\bar{q}$
are coming from
the decay of a very massive $g_{1}^{*}$, they will carry very high $p_T$. Thus,
we will see a large enhancement in the high $p_T$ jet
cross sections compared to
the usual QCD expectation.

\item {\bf High $p_T$ trijet production}
 
 $gg\rightarrow g_{1}^{*}g,\,  g_{1}^{*}\rightarrow gg, q\bar{q}$

In a very high energy hadronic collider, such as LHC, above the threshold of 
$g_{1}^{*}$ production, an onshell $g_{1}^{*}$ together with a gluon will be
produced. $g_{1}^{*}$ will decay to gg and $ q\bar{q}$ giving rise to two very
high $p_T$
jets. Thus, above this threshold, there will be anomalously large production
of three jet events with two of the jets  having very large $p_T$, and the
third with somewhat
lower $p_T$. Such events will be distinct from the usual QCD prediction.
The decay $g_{1}^{*}\rightarrow ggg$ (which
is somewhat suppressed compared to $g_{1}{*}\rightarrow gg$ decay) will
lead to four high $p_T$ jets in the final state with three of the jets
having very large $p_T$.

\item {\bf Pair productions of KK gluons and high $p_T$ four jet signals}

The relevant subprocess for for the four jet signal is

$gg\rightarrow g_{1}^{*}g_{1}^{*},\, g_{1}^{*}\rightarrow gg,\,q\bar{q}$

This will be the most important signal in our scenario at the LHC energy
($\sqrt{s}=14$ TeV). If the compactification scale for a ``merely large" 
extra dimension is in the few TeV range, the first excitation of the gluon
$g_{1}^{*}$, will be pair produced, and each one will decay dominantly into 
$q\bar{q}$, or gg. Since the quark, antiquark, or the two gluons are coming
from the decay of a very massive particle, they will have very high $p_T$,
much higher than produced in the usual QCD process. Thus, above this $g_{1}^{*}$
$g_{1}^{*}$ pair production threshold, there will be anomalous large production of
high $p_T$ four jet events. These events will be very distinct from the usual
QCD production, and will constitute the smoking gun signal for our model.

\item{\bf Top Production}

In our scenario with TeV$^{-1}$ scale extra dimensions, there is a new source for 
the production of the top quark anti-top quark pairs. The relevant subprocesses are:

$q\bar{q}\rightarrow g_{1}^{*}\rightarrow t\bar{t}$

$gg\rightarrow g_{1}^{*}g_{1}^{*}, g_{1}^{*}\rightarrow t\bar{t}$

The cross section for the top production at the hadronic colliders will be
significantly altered due to the contributions from the low lying KK states.
The first subprocess above is best explored at the upgraded Tevatron due to 
the high $q\bar{q}$ luminosity. As discussed in section 3, the $t\bar{t}$
cross section will be less than that expected from the SM below the scale of 
resonance production due to the negative interference effect. With the high
luminosity, it may be possible to study such a deviation at Run 2. The second
subprocess is best studied at the LHC energy. The four top (anti-top) production
cross section will be significantly higher than expected from QCD. As in jet 
physics, the $p_T$ distributions of the top will be significantly altered 
compared to the SM expectations.

\end{enumerate}

\section{Caveats and outlook}

We admit that we don't know how to generate or stabilize the scales
$R$ or $r$, and furthermore it is a little strange that one (or more) of the
$r$ type dimensions has a brane extending in it while the other
$r$ type dimensions don't.
We also admit that we haven't shown why it is natural to have
$M \sim 100\mu_0$; all we have really argued is that this may
be consistent with a picture involving accelerated gauge coupling
unification.
At least $R$ and $r$ correspond to
scales that already exist in physics (the cosmological constant and
the electroweak scale, respectively).

Our scenario predicts a host of new phenomena that can be tested in
upcoming high energy collider experiments. Of course these phenomena
are common to many other scenarios with TeV KK scales.
We hope to come back to the details
of these phenomena in a future work.\\[0.3in]

We are very grateful to S. Cullen and D. Smith for useful communications.
SN wishes to thank the Fermilab Theory Dept. for warm hospitality
and support during the completion of this work. The work of SN was 
supported in part by the U. S. Department of Energy Grant Number
DE-FG03-98ER41076; the work of JL by DE-AC02-76CHO3000.

\bibliographystyle{unsrt} 

\newpage
\begin{figure}
\label{run2fig}
\epsfbox[76 230 495 549]{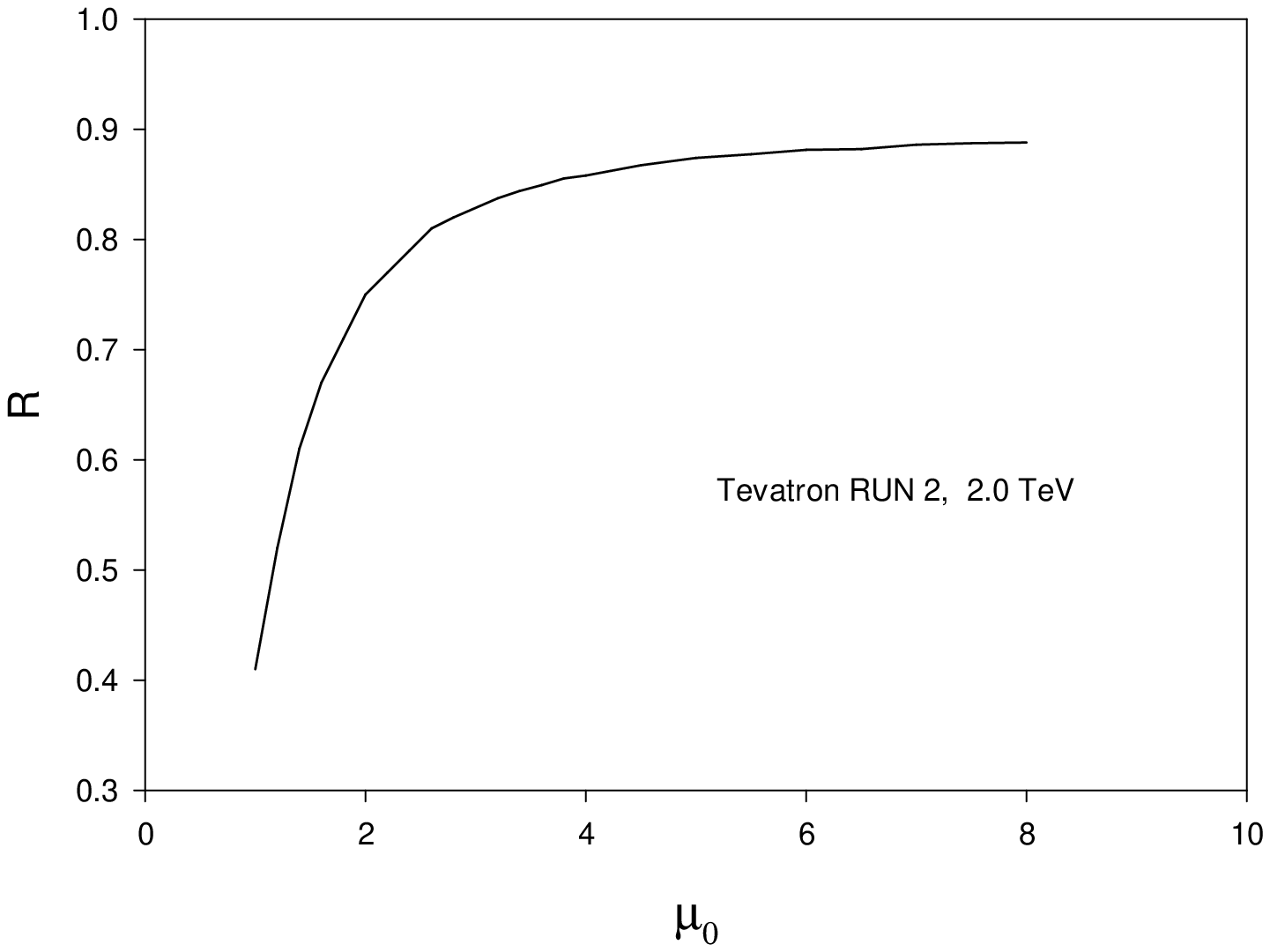}
\caption{Ratio of the cross section for our model at scale $\mu_{0}$ 
over the SM cross section for $t\bar{t}$ production.}
\end{figure}
\end{document}